\newcommand{\sys}[1]{\textit{DISCERN}}
\newcommand {\ed}[1]{{\color{black}{#1}}}
\newcommand {\camera}[1]{{\color{black}{#1}}}
\documentclass[sigconf]{acmart}

\usepackage{multicol}
\usepackage{xcolor}
\usepackage{color}

\colorlet{tableheadcolor}{gray!25} 
\colorlet{tablerowcolor}{gray!15} 
\colorlet{tablerowcolor2}{gray!12} 
\colorlet{tablerowcolor3}{gray!25} 
\colorlet{tablerowcolor4}{gray!50} 

\usepackage{tikz}
\usetikzlibrary{shapes,backgrounds}

\newcommand{\graycircle}[1]{%
\begin{tikzpicture}[baseline=-0.5ex]
    \node[circle, fill=gray!70, inner sep=2pt, text=white] {#1};  
\end{tikzpicture}%
}
\AtBeginDocument{%
  \providecommand\BibTeX{{%
    \normalfont B\kern-0.5em{\scshape i\kern-0.25em b}\kern-0.8em\TeX}}}

\copyrightyear{2024}
\acmYear{2024}
\setcopyright{rightsretained}
\acmConference[CHI '24]{Proceedings of the CHI Conference on Human Factors
in Computing Systems}{May 11--16, 2024}{Honolulu, HI, USA}
\acmBooktitle{Proceedings of the CHI Conference on Human Factors in
Computing Systems (CHI '24), May 11--16, 2024, Honolulu, HI, USA}
\acmDOI{10.1145/3613904.3642685}
\acmISBN{979-8-4007-0330-0/24/05}

\begin{document}

\title[\sys{}: Designing Decision Support Interfaces to Investigate\\the Complexities of Workplace Social Decision-Making With Line Managers]{\sys{}: Designing Decision Support Interfaces to Investigate the Complexities of Workplace Social Decision-Making With Line Managers}

\author{Pranav Khadpe}
\email{pkhadpe@cs.cmu.edu}
\affiliation{%
  \institution{Carnegie Mellon University}
  \city{Pittsburgh}
  \state{Pennsylvania}
  \country{USA}
}

\author{Lindy Le}
\email{lindle@microsoft.com}
\affiliation{%
  \institution{Microsoft}
  \city{Redmond}
  \state{Washington}
  \country{USA}
}

\author{Kate Nowak}
\email{kanowak@microsoft.com}
\affiliation{%
  \institution{Microsoft}
  \city{Redmond}
  \state{Washington}
  \country{USA}
}
\author{Shamsi Iqbal}
\email{shamsi@microsoft.com}
\affiliation{%
  \institution{Microsoft}
  \city{Redmond}
  \state{Washington}
  \country{USA}
}

\author{Jina Suh}
\email{jinsuh@microsoft.com}
\affiliation{%
  \institution{Microsoft}
  \city{Redmond}
  \state{Washington}
  \country{USA}
}

\renewcommand{\shortauthors}{Khadpe, et al.}

\begin{abstract}
\ed{Line managers form the first level of management in organizations, and must make complex decisions, while maintaining relationships with those impacted by their decisions. Amidst growing interest in technology-supported decision-making at work, their needs remain understudied. Further, most existing design knowledge for supporting social decision-making comes from domains where decision-makers are more socially detached from those they decide for. We conducted iterative design research with line managers within a technology organization, investigating decision-making practices, and opportunities for technological support. Through formative research, development of a decision-representation tool—\sys{}—and user enactments, we identify their communication and analysis needs that lack adequate support. We found they preferred tools for externalizing reasoning rather than tools that replace interpersonal interactions, and they wanted tools to support a range of intuitive and calculative decision-making. We discuss how design of social decision-making supports, especially in the workplace, can more explicitly support highly interactional social decision-making.}

\end{abstract}



\maketitle
\section{Introduction}
\ed{Within many social institutions, the task of choosing one course of action over another, falls on individual decision-makers with relevant knowledge or authority. As work, social life, and civic participation occur through increasingly digital environments, there has been growing interest, within the HCI community, in how digital tools can support this social decision-making~\cite{kawakami2022care, lee2013social, mahyar2020designing, gordon2022jury, jasim2021communityclick}, where decision-makers must reason or make predictions about the intentions, beliefs, and behavior of other people~\cite{kawakami2022care, lee2013social}. For example, a city planner might need to compare proposed public service initiatives while juggling a list of costs and benefits to many parties. Careful analysis can quickly overwhelm a decision-maker. Prior work has contributed systems that attempt to support decision-makers by providing them with the technological means to judge the values of affected stakeholders and consider the ramifications of different courses of action, till a superior course of action can be identified. From this work, three influential approaches to support social decision-making have emerged: tools that digitally elicit stakeholder perspectives~\cite{jasim2021communityclick, salganik2015wiki, desouza2017technology, boulianne2018citizen}, visual analytics systems that help decision-makers numerically analyze competing values~\cite{gratzl2013lineup, pajer2016weightlifter, carenini2004valuecharts}, and forecasting systems that help decision-makers anticipate likely outcomes~\cite{guo2019visualizing, gordon2022jury, buccinca2023aha, chouldechova2018case, tan2018investigating}. 

Do the nature of social decision-making and opportunities for technological support change when the decision-maker and those impacted by the decision must interact more closely? Even within social decision-making, HCI research so far has primarily studied decision-makers who are socially detached from the targets of their decisions, such that the decision-maker doesn't have close relationships with those they judge. This detachment is intentional in some cases (e.g., it is desirable that the decision of a content moderator not be distorted by their relationship with the targets of the decision ~\cite{pan2022comparing}), or a result of organizational scale (e.g., government officials may have limited interaction with citizen~\cite{mahyar2020designing}). However, many decision-making scenarios fall somewhere between entirely personal decisions and largely detached social decisions; decisions where formalization is not entirely absent nor very high. 

During our research studying workplace decision-making within a large technology organization, we discovered one class of social decision-makers, line managers, with decision-making practices that fall in this middle ground. Line managers---shop-floor supervisors, leaders of sales teams, managers in call centers---form the first level of management across several social institutions. Typically, they make up as much as 60\% of management ranks in most companies and supervise as much as 80\% of the workforce~\cite{fred-2011}. The datafication of most workplaces has generated excitement about developing decision-making supports at the level of line management decisions (e.g., team projects and work practices)~\cite{fitz2014predictive}. Yet, as recent work argues, little is understood about the needs of decision-makers within organizations~\cite{dimara2021unmet, oral2023information}, especially at lower levels~\cite{dimara2021unmet}. Existing research suggests that, in such settings, decision-makers may be more attuned to their relationships with those they decide for~\cite{lee2015making, blau2017exchange}. Further, the proximity to those impacted by their decisions allows decision-makers to pursue more interactional decision-making~\cite{dimara2021unmet}. In what ways should technology support social decision-making practices at this level?

In this paper, we present findings from our iterative design research that was aimed at understanding how line managers make decisions within their teams and identifying if (and how) digital tools can support it. We report on a series of design activities conducted with line managers in a large tech company, reflecting on our activities, designed artifacts, how line managers responded to our design choices, and the design opportunities they (and we) identified.

Our formative research found that line managers' decision-making practices lacked adequate support in existing tools. They described using different means to actively listen to stakeholders, resulting in (often qualitative) information that spanned both verbal and written discourse. To weigh qualitative and quantitative information within the context of different objectives, they often created makeshift structures, such as comparison tables or pros and cons lists, on spreadsheets and notepads. However, these materials were used opportunistically, leaving line managers with the choice to either use structured representations (e.g, spreadsheets)---intended to hold data---to reason about abstract objectives or instead, use more expressive materials that may be severed from the data (e.g., note-taking tools). Finally, decision-making progressed iteratively, whereby line managers made judgments by seeking inputs and understanding at intermediate points. It was often prohibitively time-consuming to create representations to anchor these intermediate discussions and record conclusions, which made it challenging for line managers to provide retrospective explanations. 

Based on our formative research, we developed a technology probe \sys{}, instantiated as an extension for Microsoft Excel. \sys{} has two main functions motivated by our formative work: (1) it supports the externalization of multi-level decisions as a tree artifact;
and (2) at each node of the tree, it provides both a visual representation---to support reasoning about objectives---and a linked tabular representation---to hold and reason about data. Finally, we conducted user enactments, asking four line managers to role-play a loosely scripted scenario in which they attempted to make a team decision---with and without \sys{}---while conferencing with a group of individuals who were role-playing as their team. We used these enactments as boundary objects, inviting line managers to reflect on how \sys{} supported or complicated their needs, to observe underlying social boundaries, and further identify opportunities for possible system designs. 

Across our studies, we found that first, despite information management challenges that result from their face-to-face elicitation practices, line managers did not want tools that would obviate their listening practices; instead, they wanted representations that would simply hold what they have heard and help them reflect on where to pay attention when eliciting subsequent inputs. Line managers often described how they used the moment of elicitation to make judgments beyond what was said. Further, even if tools could help sense opinions, they were thought to be incapable of making people feel heard: moments of personal interaction allowed the activation of pre-established trust, which was necessary for compromise. Second, we found that line managers wanted tools that would support their practice of ``qualculation''. Their decisions varied in the extent to which they were based on purely calculative conclusions or on purely qualitative intuitions, and they wanted tools that allowed this flexibility. Third, we found that they did not want to publicly and precisely quantify the importance placed on different objectives because this could threaten consensus-building efforts by making members of the team more attuned to who might gain and who might lose from each option.

These observations lend support to previous calls to look beyond data analysis and forecasting systems and include tools for interactional decision-making as a greater priority~\cite{dimara2021unmet}. Further, line managers' practice of qualculation suggests that the level of numerical analysis necessary for a decision ought to depend on the situation: what constitutes appropriate action depends on what is accountably justified in the eyes of those experiencing the situation. A sole focus on supporting highly formalized decisions and their precise mathematical modeling may overlook the many decisions within organizations that do not require that level of analysis but may still benefit from technological support. Finally, for more intimate social decision-making, our work suggests that representations intended to anchor deliberative conversations can benefit from a layer of ambiguity or fuzziness, rather than representations pinpointing values and disagreement numerically. Based on these findings across our research activities, we synthesize a set of opportunities that can guide future research and practice in designing digital tools to support social decision-making.}

\section{Related Work}
\ed{Our paper builds upon prior HCI research on supporting social decision-making. From this work, we identify, in broad strokes, three common premises underlying current tools, about how decision-making can be supported. Then, we describe how decision-making by line managers departs from previously studied domains, thus giving us an opportunity to investigate, through design, the limits of the three common premises. We suggest that our work can be interpreted as an attempt to investigate the extent to which these premises align with managers' own beliefs and practices, and therefore, the extent to which they may be applied towards developing decision support tools for line managers.}

\ed{\subsection{Current priorities in the design of social decision-making supports within HCI}
\label{priorwork}
Recent research in HCI has focused on understanding how to design technological support for social decision-making~\cite{kawakami2022care, gordon2022jury, jasim2021communitypulse, mahyar2020designing}. In domains ranging from urban planning and administration to content moderation and social work, individual decision-makers are responsible for making decisions that rest on judgements about other individuals who are then subject to the decision. For instance, in deciding whether to publish a given piece of content, content moderators in an online community make judgements about the author's intent and potential harms to other members. Because they rest on inferences about other people's beliefs, and because the impacts are felt by others, these decision-making tasks have been referred to as ``social decision-making'' in prior work~\cite{kawakami2022care, lee2013social}.

Why is social decision-making challenging? In their book \textit{Decision Analysis for Healthcare Managers}~\cite{alemi2007decision}, Alemi and Gustafson suggest a decision has the following five components:
``(1) multiple alternatives or options are available; (2) each alternative leads to a series of consequences; (3) the decision-maker is uncertain about what might happen; (4) the decision-maker has different preferences about outcomes associated with various consequences; (5) a decision involves choosing among uncertain outcomes with different values.'' Decisions, then, can be challenging when decision-makers are unsure about what criteria they ought to consider (\textit{unclear values}) or uncertain about how the options fare along these criteria (\textit{uncertain outcomes}). In social decisions, unclear values and uncertain outcomes are further amplified by stakeholders' value conflicts and uncertainty in how stakeholders would evaluate outcomes~\cite{chen2023judgment, gordon2021disagreement}. Furthermore, as the last point in the the definition suggests, an ``optimal'' alternative does not exist at the outset; rather, the superiority of alternatives depends on
the decision-maker's preferences. For a decision-maker, gathering relevant perspectives, clarifying values, and making confident predictions can quickly get overwhelming. With these challenges in mind, HCI researchers have been interested in social decision-making supports---both designing them and critically studying their uptake---because digital tools provide the means to accelerate the elicitation of stakeholder perspectives~\cite{levy2021algorithms, mahyar2020designing, kawakami2022care} and facilitate the complex analysis that real-world decision-making entails~\cite{gratzl2013lineup, pajer2016weightlifter, carenini2004valuecharts}. 

Suchman argues that ``every human tool relies upon, and reifies, some underlying conception of the activity that it is designed to support''~\cite{suchman1987plans}. How do current tools conceive social decision-making? From our review of prior HCI and CSCW research, we identify, in broad strokes, three premises for how systems currently support social decision-making. This overview is not intended to be a complete discussion of all decision support systems within HCI, and not every system illustrates every one of these premises. Rather, we intend to highlight some common ways in which social decision-making has been supported in prior research; stated either explicitly in the paper or implicit in the approach taken. We introduce these premises as briefly as possible before discussing how they are realized in specific systems. The premises are as follows: (1) Stakeholders' perspectives often form the basis of a decision-maker's value clarification and outcome forecasting activities so digital tools for eliciting stakeholder perspectives as data can make decision-makers more efficient. (2) Quantifying values, however much they resist quantification, is unavoidable when trying to rank courses of action so numerical representation of values can support decision-makers in efficiently clarifying their values. (3) Uncertain outcomes make it hard to compare options so outcome forecasting systems can support decision-making. Next, we describe how they are instantiated in specific systems:

\paragraph{Digitally eliciting stakeholder perspectives:} Digitally distributed polls and surveys are already a common way to elicit stakeholder opinions~\cite{desouza2017technology, boulianne2018citizen}, and researchers continue to explore technologies that accelerate data gathering efforts~\cite{jasim2021communityclick, salganik2015wiki}. In some cases, digitally eliciting perspectives is a response to the challenges of large scale participation~\cite{jasim2021communityclick, mahyar2020designing}. For example, online communities have increasingly aggregated individual beliefs via digital means~\cite{georgakopoulos2018convolutional, van2018challenges, zhou2018fake, zilouchian2015procid}, for content moderation. In other cases, digital representation of perspectives is often viewed as necessary precursor to being able to quantify opinions and pinpoint conflicting values~\cite{liu2018consensus}. This has led researchers to explore how information elicitation strategies can structure stakeholder responses to maximize insights for the decision-maker~\cite{kriplean2012supporting, lyons2014exploring}. Together, digital elicitation advances some notion of efficiency.   

\paragraph{Numerically representing values:} To help decision-makers consider and articulate their values, researchers have contributed visualizations and interaction techniques that support numerical modeling of values~\cite{gratzl2013lineup, pajer2016weightlifter, carenini2004valuecharts}. These are often used in conjuction with digital elicitation methods: they impose structure on gathered data and at the same time, help decision-makers identify where more perspectives or data ought to be gathered. The idea of enforcing numerical representation as a way to advance decision-making is also found in HCI work on group judgments and problem solving~\cite{alonso2007using, alonso2007visualizing, liu2018consensus}. Much of this work shares an implicit or explicit commitment to the idea that quantifying our values, however much they resist quantification, is unavoidable when trying to rank courses of action~\cite{alemi2007decision}. 

\paragraph{Forecasting systems:} Researchers have also used digitally aggregated data to create forecasting systems to help decision-makers anticipate likely outcomes based on what values are prioritized~\cite{gordon2022jury, buccinca2023aha}. For decisions that are even more formalized, exploratory interfaces have been replaced with an algorithmic assessment of outcomes that are based on predetermined values~\cite{chouldechova2018case, tan2018investigating, kawakami2022improving}. These systems are not without criticism---they can inspire overconfidence and even obscure the values they reify~\cite{kawakami2022improving}, which has motivated further work into designing interfaces and explanation mechanisms that might support more critical use~\cite{kawakami2023training}.

Although our immediate focus is on studying and designing for decision-making by line managers, we suggest our work can be seen as replicating and testing the limits of these underlying premises as applied to the decision-making practices of line managers, an understudied social decision-maker~\cite{dimara2021unmet}. We next describe how studying line managers provides us with an opportunity to evaluate the premises reified in prior work under a new context. It allows us to explore the implications of their success and failure for the design of support for social decision-making.
}

\ed{\subsection{Line managers as decision-makers: a new sphere for social decision-making support in HCI}
By and large, HCI research on the design and deployment of social decision-making supports has focused on decisions marked by a level of social detachment, where the decision-maker does not have a close, ongoing relationship with the people they make judgements about. In some settings, this separation is by design: It is intended to prevent decisions from being distorted by a concern for one’s own reputation. For instance, the expert moderator does not seek the social approval of those who post offending content~\cite{pan2022comparing}. In other settings, the detachment is a consequence of the social organization's scale or communication channels. For example, government officials may have limited interaction with individual citizens impacted by their decisions~\cite{mahyar2020designing}, and online moderators may perceive themselves to be psychologically distant from content authors~\cite{hwang2021people}. However, many decision-making scenarios fall somewhere between entirely personal decisions and largely detached social decisions. These are decisions where formalization is not entirely absent nor very high, and where the decision-maker seeks social approval from the individuals for whom they are making decisions. 

In this paper, we study decision-making practices of one class of decision-makers with social decision-making practices that fall in this middle ground---line managers. Even though they make up a large portion of decision-makers in any organization, not much is understood about their practices and needs: prior work argues how research and broader discourse tends to focus only on the needs of data analysts or senior executives~\cite{dimara2021unmet}. Additionally, the datafication of most workplaces has generated excitement about developing decision-making supports at the level of line management decisions (e.g., team projects and work practices), yet little is understood about what kinds of support are appropriate~\cite{dimara2021unmet}. 

When the decision-maker has close, ongoing relationships with the people they decide for, they may be more concerned about how their decisions will affect those relationships~\cite{lee2015making, blau2017exchange}. At the same time, relational concerns also influence stakeholders in how they express their preferences~\cite{hackman1974interventions, myers1976group}. Furthermore, unlike decision-making in more detached settings where stakeholders' viewpoints are gathered at discrete points in time, decision-making in more intimate settings can be more interactional and deliberative~\cite{platt2022network}. Existing research suggests that the psychological distance between decision-makers and their stakeholders can affect decision-making activities, with greater detachment promoting more dehumanizing approaches~\cite{lee2015making, blau2017exchange}. Our study of decision-making by line managers, therefore, extends HCI research on social decision-making by contributing an understanding of how existing supports succeed and break in more intimate decision-making settings, while revealing new opportunities for future research and design.}

\section{Research and Organizational Context}
This research was conducted within a large multinational technology organization producing enterprise and consumer products. 
\ed{Our research focused on the decision-making practices of line-managers within their teams}. Like many modern workplaces, managers and teams could access several business-related data sources. For software development teams, a common data source was their developer operations platform, tracking application lifecycle activity. Like many information workplaces, workplace analytics was also available to all employees through commercially available technology, Viva Insights\footnote{https://www.microsoft.com/en-us/microsoft-viva/insights}. The software passively collects metadata about employees' activities across enterprise communication software. People could access summaries and interpretations of their own data (e.g. individual work patterns). Line managers had access to aggregated data about their teams (e.g. average hours worked overtime). Like most workplaces, the organization also incorporated institutionalized feedback methods, including performance reviews and an internal feedback tool. Together, these served as line managers' routine sources of quantitative and qualitative information. Our research activities, which we describe next, were motivated by an attempt to understand how line managers practice decision-making and what (if any) technological support they need.

\ed{\paragraph{Ethical considerations:} Our work most directly relates to research that aims to influence the \textit{process} of decision-making, such as work that improves decision-makers' efficiency and confidence~\cite{guo2019visualizing, pajer2016weightlifter}. For this reason, the sole subjects of our research are the line managers. Our work does not aim to evaluate the values---organizational or personal---that line managers prioritize in their decisions. Therefore, we do not claim that our work improves decision-making \textit{outcomes} along normative criteria such as fairness, though we recognize this is an important line of HCI research. Reflecting on our inquiry and the values it promotes, we believe that when decision-making activities are externalized in decision support tools, they are given material presence, leaving an audit trail where there might previously have been none. Moreover, line managers exist within a monitoring structure; they are themselves subject to the oversight and evaluations of higher-level executives. Decision support tools, can support peers and supervisors in monitoring for potential misuse of such tools for personal gain at the expense of others. Additionally, we find that line managers currently lack the means to capture and articulate their decision-making rationale to their teams. While decision support systems do not enforce full transparency, they create the potential records, paving the way for accountability and contestability~\cite{holten2021can}. Lastly, decision support tools can normatively encourage decision-makers to expicitly articulate what they value, which has the potential to reduce deviations from stated values~\cite{mergler2008making}. Finally, line managers are both managers and managed. As the first line of organizational decision-makers, they provide a unique perspective into the challenges they face in making decisions while considering their own concerns, as well as those of their team members, regarding the role that data should play in workplace decision-making. While our focus on line managers means our work is a limited representation of workplace decision-making, it begins to paint a richer picture of line managers as stakeholders in determining futures of workplace decision-making. Thus, our work with line managers takes the position that decision support systems have the potential to make otherwise inscrutable decision-making more legible, which supports greater regulation from both the decision-maker and other organizational actors, though we recognize that this position comes with its own limitations.}

\subsection{Participants}
\ed{Our research involved $59$ line managers (hereafter referred to simply as managers), working in the United States, across three activities.} All activities were approved by the lead institution's IRB:
\begin{itemize}
    \item \textit{Survey:} To understand managers' current practices of gathering, organizing, and analyzing decision-making information and what tools they desired, we conducted an open-ended survey with $57$ participants. Participants were told that for every response, we would donate \$$2$ to Theirworld\footnote{https://theirworld.org/}, a global chidren's charity.
    \item \textit{Semi-structured interviews:} For additional context, we interviewed $11$ participants from the survey. \ed{Before these interviews, we used survey findings in conjuction with the three premises from prior work to generate design concepts and presented these to interviewees to further identify opportunities for technological support}. Participants were compensated \$$50$ for their time.
    \item \textit{User enactments:} Based findings across the surveys and interviews, we developed a technology probe, \sys{}. Then, we invited four managers to participate in user enactments, role-playing a decision-making scenario that utilized our probe to assess the viability of our ideas and generate new questions around designing decision-support tools. Two managers had participated in the survey (but not the interview), and two had not participated in any previous activity. This took place across two sessions and participants were compensated \$$75$.
\end{itemize}

\begin{table*}[t]
\sffamily
\vspace{-0.5em}
\label{table:demographics}
  \centering
    \footnotesize
    \begin{minipage}[t]{0.75\textwidth}
    \begin{tabular}{lrr}
 \textbf{Response} & \textbf{Count} & \textbf{Percentage} \\
 \cmidrule(lr){1-1}\cmidrule(lr){2-2}\cmidrule(lr){3-3}

\multicolumn{3}{l}{{{\textbf{Gender}}}}\\
Male                 & 44    & 74.58\%    \\
Female               & 11    & 18.64\%    \\
Woman                & 1     & 1.69\%     \\
Prefer not to say    & 3     & 5.08\%     \\
\hline
\multicolumn{3}{l}{{{\textbf{Age}}}}\\
26-35 years old      & 5     & 8.47\%     \\
36-45 years old      & 16    & 27.12\%    \\
46-55 years old      & 23    & 38.98\%    \\
56-65 years old      & 12    & 20.34\%    \\
Prefer not to say    & 3     & 5.08\%     \\
\hline
\multicolumn{3}{l}{{{\textbf{Tenure}}}}\\
1-3 years             & 5  & 8.47\%  \\
3-5 years             & 5  & 8.47\%  \\
5-7 years             & 6  & 10.17\% \\
\textgreater{}7 years & 42 & 71.19\% \\
Prefer not to say     & 1  & 1.69\% \\
\bottomrule
\end{tabular}
\hfill
\begin{tabular}{lrr}
\textbf{Response} & \textbf{Count} & \textbf{Percentage} \\
 \cmidrule(lr){1-1}\cmidrule(lr){2-2}\cmidrule(lr){3-3}
\multicolumn{3}{l}{\textbf{Total Reports (both direct and indirect)}}\\
5 or less                         & 19    & 32.2\%     \\
6-10            & 20    & 33.9\%     \\
11-20           & 9     & 15.3\%     \\
21-30           & 7     & 11.9\%     \\
31-50           & 2     & 3.4\%      \\
More than 50                         & 1     & 1.7\%      \\
Prefer not to say                      & 1     & 1.7\%      \\
\hline
\multicolumn{3}{l}{\textbf{Profession}}\\
Engineering and Software   Development & 41    & 69.49\%    \\
Management                             & 6     & 10.17\%    \\
Business Operations                    & 3     & 5.08\%     \\
Research                               & 3     & 5.08\%     \\
Sales and Marketing                    & 1     & 1.69\%     \\
Design and Creative                    & 1     & 1.69\%     \\
Compliance and Ethics                  & 1     & 1.69\%     \\
Other                                  & 2     & 3.39\%     \\
Prefer not to say                      & 1     & 1.69\%    \\
\bottomrule
\end{tabular}
\end{minipage}
\caption{Distribution of participants in the study, along with count and percentage of participants in the pool. The data is aggregated to protect participant privacy.}

\label{demographics}
\end{table*}

Table \ref{demographics} shows demographics for the $58$ participants (one declined) who opted to share their demographics, aggregated to protect participant privacy. \ed{It also shows the size of the teams (Total Reports) of managers in our study. The distribution of team sizes---a majority comprising less than 10 members with a drop off subsequently---are consistent with those typically reported for line managers in companies with more than 250 workers~\cite{urquhart-2022}.} 

\section{Formative Studies}
\label{formative}
To guide our exploration of how digital tools might support organizational decision-making, we conducted formative research with managers. Our goals were to: (1) understand how managers currently gather, organize, and analyze decision-making information, and (2) identify opportunities where digital tools can support current practices. Our formative research progressed in two stages: a survey followed by semi-structured interviews. 

\subsection{Survey}
The survey asked participants about current team decisions---what information they anticipate gathering, what tools they typically use to analyze their data, and what tools they would desire. The survey was distributed within the organization via email containing the consent form. We received 57 complete responses. To analyze our data, we used an affinity diagramming approach, grouping responses into categories of tools and information used and the kinds of support managers desired. We drew three main conclusions, which guided our semi-structured interviews (percentages indicate the number of respondents mentioning a specific category; since the surveys were open-ended, they don't sum to 100\%):

\textit{Managers seek decision-making input from multiple sources.} Participants primarily solicited new information from relevant stakeholders (71\%), through informal conversations (65\%), such as  1-1 meetings or team meetings (53\%). Fewer participants solicited information through digital questionnaires (13\%). Participants described using institutionalized feedback channels (29\%) and conversation histories (22\%) to retrieve retrospective decision-making information. Application lifecycle management tools (42\%) were valuable sources of work content; however, managers often described augmenting these with personal observations about work quality and impact (42\%). Few managers described gathering data from workplace analytics tools (5\%). Finally, several managers described how personal observations of how team members operated were an important decision-making input source (55\%).

\textit{Managers use an assemblage of tools.} Participants most commonly mentioned using note-taking tools (e.g. Word, OneNote) to organize their thoughts and observations (33\%). Similarly, many participants mentioned using spreadsheets to record information as well as conduct informal analyses (25\%). Some reported using dedicated data visualization and business analytics tools (12\%). 5\% of participants reported that they didn't use any particular tool or made decisions offline.

\textit{Managers most commonly wanted tools to listen.} Most commonly participants described wanting tools that would allow them to more immediately receive input and feedback from their team (17\%). Some participants (7\%) explicitly expressed a preference for tools that do not intrude on team members' privacy. Another 15\% were interested in tools that could estimate a measure that was personally relevant. Finally, 12\% were interested in tools to reflect on and plan team decisions.

\subsection{Semi-structured Interviews}
\begin{figure*}[t]
    \centering
    \includegraphics[width=\textwidth]{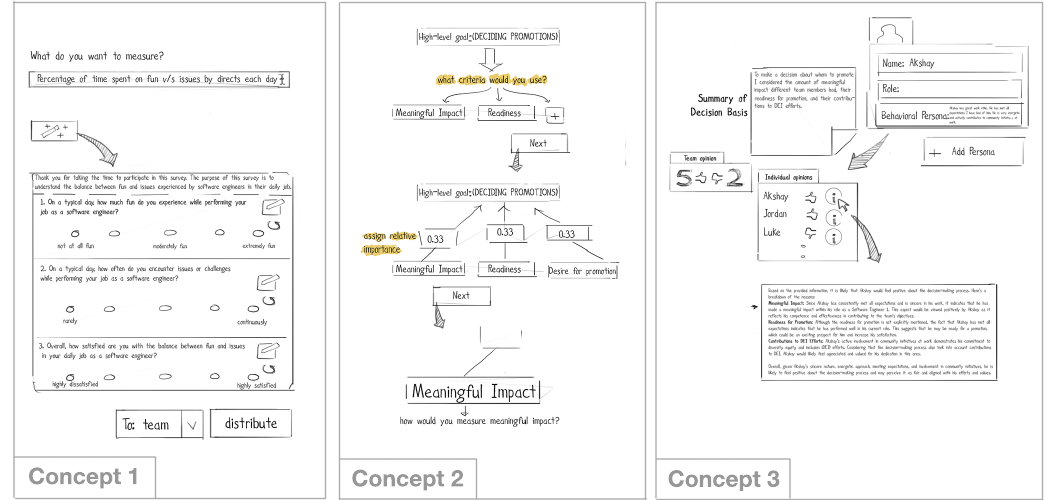}
    \caption{Concepts used in the interview to further understand managers' needs: Concept 1 is a tool that would automatically generate and administer short surveys to provide managers' near real-time feedback while preserving privacy. Concept 2 is a visual environment to author multi-attribute value representations as an environment to reflect and plan. Concept 3 is a tool that would simulate stakeholder opinions as a way to receive real-time proxies of team members' feedback when they may not be accessible.}
    \Description{
    “Concept 1 asks ‘What do you want to measure?’ above a textbox with ‘Percentage of time spent on fun v/s issues by directs each day’ typed in. Underneath is a magic wand button that generates a multiple-choice survey whose purpose is to understand the balance between fun and issues experienced by software engineers in their daily job. Question one says, ‘On a typical day, how much fun do you experience while performing your job as a software engineer?’ with answers options from ‘not at all fun’ to ‘moderately fun’ to ‘extremely fun.’ Question two says, ‘On a typical day, how often do you encounter issues or challenges while performing your job as a software engineer?’ with options ranging from ‘rarely’ to ‘continuously.’ Question three says, ‘Overall, how satisfied are you with the balance between fun and issues in your daily job as a software engineer?’ with options ranging from ‘highly dissatisfied’ to ‘highly satisfied.’ Below the survey is a drop-down box with ‘To: team’ written inside and a ‘distribute’ button to the right of it. Concept 2 consists of iterative schema graphs. The topmost one is headed by ‘High-level goal: [DECIDING PROMOTIONS]’ which leads to the question ‘What criteria would you use?’ That question is further decomposed into attributes such as ‘Meaningful Impact,’ ‘Readiness,’ and a ‘+’ button to add other attributes.” A ‘Next’ button points to a similar schema graph headed by ‘High-level goal: [DECIDING PROMOTIONS].’ This time, three attributes (‘Meaningful Impact,’ ‘Readiness,’ and ‘Desire for Promotion’ are linked to ‘High-level goal: [DECIDING PROMOTIONS].’ Each attribute is assigned a relative importance of 0.33. A ‘Next’ button under ‘Meaningful Impact’ points to the question ‘How would you measure meaningful impact?’ Concept 3 contains a persona box with three components: ‘Name’, ‘Role’, and ‘Behavioral Persona’. As an example, ‘Akshay’ is written in for ‘Name’ and ‘Akshay has good work ethic. He has met all expectations I have had of him. He is very energetic and actively contributes to community initiatives at work.’ for ‘Behavioral Persona.’ Below is a box labelled ‘+ Add Persona.’ Akshay’s persona box feeds into a box labelled ‘Individual opinions,’ which consists of three names: ‘Akshay,’ ‘Jordan,’ and ‘Luke.’ Beside ‘Akshay’ and ‘Jordan’ is a thumbs up. Beside ‘Luke’ is a thumbs down. Each thumbs up or down is followed by an information button. To the left of the ‘Individual opinions’ box is a ‘Team opinion’ box with 5 thumbs up and 2 thumbs down. Above the ‘Individual opinions’ box is a ‘Summary of Decision Basis’ sticky note, filled with: ‘To make a decision about whom to promote, I considered the amount of meaningful impact different team members had, their readiness for promotion, and their contributions to DEI efforts.’ Clicking on Akshay’s information button reveals a box filled with: ‘Based on the provided information, it is likely that Akshay would feel positive about the decision-making process. Here’s a breakdown of the reasons: Meaningful Impact: Since Akshay has consistently met all expectations and is sincere in his work, it indicates that he has made a meaningful impact within his role as a Software Engineer I. This aspect would be viewed positively by Akshay as it reflects his competence and effectiveness in contributing to the team’s objectives. Readiness for Promotion: Although the readiness for promotion is not explicitly mentioned, the fact that Akshay has met all of expectations indicates that he has performed well in his current role. This suggests that he may be ready for a promotion, which could be an exciting prospect for him and increase his satisfaction. Contributions to DEI Efforts: Akshay’s active involvement in community initiatives at work demonstrates his commitment to diversity, equity, and inclusion (DEI) efforts. Considering that the decision-making process also took into account contributions to SEI, Akshay would likely feel appreciated and valued for his dedication in this area. Overall, given Akshay’s sincere actions, energetic approach, meeting expectations, and involvement in community initiatives, he is likely to feel positive about the decision-making process and may perceive it as fair and aligned with his efforts and values.’"
    }
    \label{concepts}
\end{figure*}
To further understand managers' decision-making practice and identify opportunities for technological support, we next conducted semi-structured interviews with 11 managers, selected from a list of survey respondents who agreed to be contacted for future studies. Interview questions focused on the decisions participants had mentioned in the survey, asking participants how they sourced, organized, and analyzed information relevant to a decision. To understand managers' preferences for tool support, we generated three design concepts: Concept 1 is a tool that automatically generates and administers short surveys to provide near real-time feedback while preserving privacy. Concept 2 is a visual environment to author multi-attribute objectives, reflect, and plan. Concept 3 is a tool that simulates stakeholder opinions as real-time proxies of team members' feedback. These concepts were generated to address needs mentioned in the formative survey (specifically managers' desire for support with listening (Concept 1 and Concept 3) and planning (Concept 2)) and \ed{were inspired by the three premises we identified in prior work (Section \ref{priorwork})}. \ed{Specifically, Concept 1 is a tool that digitally elicits stakeholder perspectives, Concept 2 is a tool that supports numerical representation of values, and Concept 3 is a forecasting system}. Figure \ref{concepts} shows sketches of the concepts shown to managers. \ed{We used the concepts to see how strongly participants identified with the issues they tried to address, and whether they felt the underlying assumptions were appropriate}. We asked interviewees what they liked and disliked, and we encouraged them to articulate alternate visions of technologies they would desire. Interviews were conducted by the first author, remotely via video call, and lasted 60 minutes. To analyze our data, we organized our observations using an affinity diagram approach. We organize our findings in two sections. The first section describes how participants currently \textit{practice} decision-making and the second describes the \textit{preferences} our participants expressed for technological support. Due to the low number of participants in our interviews and their shared organizational context, these findings should be regarded as indicative.

\subsubsection{\textbf{The practice of managerial decision-making}}
\paragraph{(1) Interactions with stakeholders are deemed necessary to enrich raw data with knowledge and preferences.}
All managers described engaging in information-seeking practices that draw on the knowledge of different stakeholders. Through these processes, managers contextualize and enrich raw data. P8 noted, \textit{``The toughest part of a lot of the issues is figuring out what the real issues are, which are usually only figured out through person-to-person discussions.''} P7 expressed a similar sentiment: \textit{``Telemetry can tell you a story. But, you know, how often have you seen a story in the newspaper that's backed by data but is misleading?''} Early-stage conversations with different stakeholders were viewed as unavoidable, because they were the primary means to understand and reconcile preferences. All managers described how, after identifying that a decision needs to be made, they often used team meetings, 1-1s, and surveys, to understand individual team members' preferences (e.g. project assignments). Similarly, managers often met with peers and partner teams to understand the objectives of other actors. P2 illustrated these early-stage conversations: 
\begin{quote}
    \textit{What should we do with this thing? Uh, my next question might be, I guess, to seek more information, which would be, I might say, ``What is the desired outcome like?'' Or I might say it as, ``Is the desired outcome X or Y or something else?''}
\end{quote}

\paragraph{(2) Managers create visual representations in both early- and late-stages to structure their decisions, but the search for good representations is constrained by the materiality of their tools.}
Most managers (7/11) externalized their thought processes as visual cognitive maps at different stages of their decision-making processes. The specific tools used by different managers included spreadsheets, text documents, and whiteboards. In the early stages, managers created visual representations to clarify ill-expressed objectives and understand the problem: 
\begin{quote}
    \textit{I wrote down everything that I was trying to follow up on on a whiteboard. And then tried to mark them as you know, ``Is this urgent? Is this something we need to think about eventually?'' And then if I can actually draw that all out, it doesn't even take that long. If I do that, then it becomes clear.} (P2)
\end{quote}

In later stages, managers created representations, such as comparison tables, to juxtapose and discriminate between alternatives. Many late-stage representations were structured as tables and directly populated with evaluative data. For this reason, managers often created them in spreadsheets, to easily import data from different sources.

However, many managers complained that tool switching was prohibitively time-consuming because the materiality of each tool only permitted a certain level of analysis:\textit{ ``You can either build that tree or have the data''} (P6). Some managers manually copied over early-stage visual representations into spreadsheets during later stages. Due to the effort involved, some managers attempted to externalize their early-stage maps within spreadsheets but found the result \textit{``clunky''} (P11). We interpreted this as a mismatch between the high-level nature of early-stage reasoning and the low-level granularity of spreadsheets. As one participant described, \textit{``I use tools of opportunity right now because I haven't found a tool that does what I want''} (P6).
\paragraph{(3) Externalizations of decisions also help mobilize, record, and explain.}
\label{repforinf}
Several managers used externalizations they had created---either by sharing these directly or using them for personal reference during conversations---to mobilize information-seeking and consensus-building activities. Managers used structured externalizations to coordinate input and represent the current state of the decision: \textit{``I got feedback from each employee and basically built a matrix that then, sort of, pushed us in that direction''} (P3). Sometimes managers kept their externalizations private to invite perspectives and contrasting opinions without overwhelming stakeholders. For instance, when mobilizing information-seeking, managers asked, \textit{``Give me pros and cons''} (P2), and \textit{``These are our options. What do you guys think?''} (P9). On the other hand, when establishing consensus on decision criteria, managers explained, \textit{``You think the problem is this. I think the problem is that''} (P2), and \textit{``That's why we created the spreadsheet is so that we all buy into the process''} (P11).

Externalized representation also served as a decision-making record, used retrospectively to explain decisions: \textit{``It creates transparency which not everybody will be happy with, but it's there, right? And it's defensible. It's in a position where you know you can say, `No, no, like I followed this step'''} (P10). However, managers often kept the full externalization private: (1) to preserve the confidential conditions under which information was initially shared (e.g.,\textit{``sharing information from a manager-only meeting that is appropriate to share''} (P4)); (2) to preserve a level of ambiguity  deemed necessary for face management (e.g., \textit{``I mean, I wouldn't necessarily say, `Ohh, well this person had more experience. So they got [the project].' ... I mean, they could figure that out''} (P11), and \textit{``There's a balance of externalizing it entirely, maybe sharing a rubric''} (P10)); 
and (3) to present only critical information and avoid overwhelm: 
\begin{quote}
    \textit{I would use [spreadsheet] more personally just so that I can work my way to clarity out of a couple of close alternatives and help me. Like when I was articulating to others, I would be probably listing a couple, you know, 2 to 3 alternatives with pros and cons more likely.} (P8)
\end{quote}

When externalized representations did not exist or contain sufficient context, managers often felt less confident about their decisions: \textit{``I would feel more comfortable if I had others, you know, backing me up on the methodology, backing me up on the data, backing me up on X, Y \& Z, on why I made that decision''} (P1). P6 mentioned: 
\begin{quote}
    \textit{I find that when I just do this in-my-head exercise, that's fine for me to get to a point, but if I have to go and tell that story to somebody else, my thoughts might go a slightly different path or a completely different path.... If you do that too often, there's confusion everywhere, including inside myself. So writing it down helps me to organize the thoughts and make sure that I have a repeatable thought process.}
\end{quote}

\paragraph{(4) Decisions progress iteratively whereby decisions are decomposed and consensus is sought at intermediate points.} All managers reported engaging in some level of decomposition. P1 tried to measure early consensus on proposed courses of action: 
\begin{quote}
    \textit{Sometimes like we'll just ideate and float things by at a very surface level just to gauge---oh, yeah, before I, before we go deeper into a whole bunch of analysis and evaluation criteria and---all this, all this stuff that has to happen ... just to get a gut check, like a thumbnail sketch.}
\end{quote}
On the other hand, P8 and P2 attempted to find early consensus on the nature of the problem: \textit{``You know, hey, are we all agreed that we have this issue? And then what do you guys think are ways to address the issue?''} (P8). In both cases, however, the decision-making situation was decomposed into more tractable parts to focus analysis and consensus-building efforts. As P9 mentioned, \textit{``the flow affects how people perceive it.''}

\subsubsection{\textbf{Preferences for technological support}}
\paragraph{(1) Tools to guide, not replace face-to-face elicitation practices.}
Even though significant decision-making effort was spent digitizing and organizing verbal exchanges with stakeholders, managers consistently rejected the idea of tools that would digitally source this input. For example, one manager responded strongly to Concept 1 stating, \textit{``Why are we trying to minimize talking to the people that these decisions affect?''} (P9). Our interviews revealed two reasons for this. First, when sourcing judgments from their team, managers described enacting conversational scripts face-to-face to make judgments beyond what was said. This included judgments about stakeholders' confidence and flexibility, when expressing their knowledge and preferences, respectively. Second, managers believed that stakeholders providing information digitally wouldn't feel heard. Face-to-face conversations created an interstice where trust was both cultivated and demonstrated. Transparency (e.g.,\textit{``Is this something that's gonna cause problems or get in the way of your future work? Or is it something that you can accept?''} (P7)) and accountability (e.g, \textit{``If this doesn't work out then we'll do a retrospective and you know, you can say, `I told you so'''} (P7)) enabled flexibility, making it possible for initially irreconcilable preferences to evolve into a unified plan. Managers' comments in response to Concept 1 suggested that they would prefer if tools guided, rather than digitized, information-seeking. For instance, managers commented they would be more likely to use a tool if it helped them identify the \textit{``right''} questions \textit{``about what you might go do next''} (P10) that they could then ask in person: \textit{``I can use it in a team meeting rather than sending it as an electronic survey. I just use those questions''} (P6). 

\paragraph{(2) Representations to reason about higher level objectives while also capturing the source of individual judgments.}
In response to Concept 2, many managers noted how such a tool would help them understand and reason about different objectives involved (e.g., \textit{``I like the notion of forcing someone to articulate what's going into it.''} (P2)), especially in the early-stages: 
\begin{quote}
    \textit{I would use it more in the first iteration. So I have an opportunity to give a broad set and then start narrowing it down as I start putting values on it and kind of like figuring out what is really important for the decision.}~(P6)
\end{quote}
Some managers emphatically requested the ability to easily manipulate visual representations: 
\begin{quote}
    \textit{If there was a way to input it so you could sort of visually see, oh yeah, this one's a little bit smaller than this one. Like you have these bubbles and it's like, how do they interact with each other? That's all verbal right now. So that would be amazing to find a way to make it a little bit more controlled.} (P4)
\end{quote} 

Managers (P1, P3, P6, P11) mentioned how decisions externalized within such a tool could be especially helpful when they attempted to explain their decision to their team, peers, or own managers: 
\begin{quote}
    \textit{If you were in the evaluation criteria and you were going through this logic tree and you were collaborating around it, let's say; and maybe you're sharing with other people and getting their thoughts and input and ideas and feedback; and validating it more before you made that decision, arguably to, in my mind that would make a more meaningful decision.} (P1)
\end{quote} 
Many managers wanted to be able to link these high-level objectives to evidence: \textit{``It would be really great if I could attach data, right?''} (P6). P5 mentioned how it would be useful to \textit{``drill through those into like a kind of subcontext,''} presenting their vision: \textit{``You can kind of see like, yes, I'm using these things to make a decision, and here are their various levels of importance to me, right? And here are some sub-thoughts if you really need that stuff.''} As P1 summarized, \textit{``Sometimes you want that 100,000-foot view and not the 1000-foot view, and sometimes it's the opposite, right.''}

\paragraph{(3) Tools for ``qualculation''.}
Even though managers were receptive to tools to help reason about higher-level objectives, 8 of the 11 managers disliked Concept 2's numerical approach. Our numerical representations were inspired by the decision analysis practice of numerically modeling values and objectives~\cite{watson1987decision}. Prior work has demonstrated the effectiveness of this approach for preference evaluation, decision-making, and consensus-building~\cite{carenini2004valuecharts, watson1987decision}. However, this often evoked responses such as, \textit{``I'm not a numbers person. I'm more of like a color person''} (P1), or \textit{``I think there's just something about the numbers that's getting to me''} (P2). P5 wondered if it was \textit{``subtle''}  enough. While P11 appreciated having numerical weights, they still asked to be able to overwrite some of the \textit{``logic.''} We noticed some managers differentiated between representations for calculation and qualitative reasoning. Looking at Concept 1, P9 asked, \textit{``When I put in everything that I want and I hit calculate, what do I get?''} In this context, P8 mentioned how they used spreadsheets when there were \textit{``calculable metrics because otherwise, it's a visualization, not a calculator of better/best kind of thing.''} This suggests to us that some managers would prefer representations that enable more visual or non-numerical expressions when reasoning about higher-level objectives. Further, providing representations for subjective reasoning about objectives without implying calculation may align better with managers' current mental models. More generally, we conclude that some decisions lie in a middle ground between purely subjective, qualitative judgments at one end and purely quantitative, calculative work at the other. To convey this essence of managers' decision-making, we borrow the term ``qualculation'' as extended by Cochoy in sociological analysis of decision-making~\cite{cochoy2008calculation, cochoy2014sociologie}. We use it in this paper to mean ``attempts at rational judgement when calculation is not directly possible without first making subjective judgements.'' Finally, because some managers prefaced their feedback with context, e.g. promotion decisions for Concept 2, we note that the way in which ``qualculation'' is performed may vary across decisions. 
\paragraph{(4) Reflection prompts.}
Managers primarily saw the value of Concept 3 as a tool for reflection, however, some were hesitant to accept the form presented. Managers expressed how they would value reflection aids that present an \textit{``alternative point of view''} (P1) or \textit{``played devil's advocate''} (P2). However, some managers disliked the idea of creating a digital proxy of their team members to prompt this reflection: \textit{``It's weird that you could get an approximation of somebody's opinion without talking to them''} (P8). These responses suggest that reflection aids could be undesirable if they threatened the value managers placed on direct communication with those impacted by their decisions. Instead, managers proposed ideas of having an unattributed \textit{``contrarian viewpoint''} (P2) and reflection prompts informed by the organization's management training resources (P10).

\subsection{Reflections on Formative Studies}
Based on our synthesized findings, we reflect on managers' decision-making practices and how they may be supported by digital tools. Note that as with all design research, the following reflections are proposals of what the world ``could be'' as opposed to scientific descriptions of what the world ``is''~\cite{zimmerman2014research}.

First, managers' reactions to our concepts suggest a desire for tools that support, but not obviate, active listening. Rather than automating listening, managers want help identifying focal points in the conversation. In their own practice, managers often craft makeshift representations to identify where to direct their attention. They also gravitated towards the idea of externalized representations as a way to guide information-seeking activities. Therefore, we decided to narrow our exploration to representation materials that would allow managers to externalize and reason about their decisions. 

Second, managers expressed interest in being able to represent decisions at different levels. Participants valued abstract representations for understanding and expressing objectives. However, they also wanted the ability to explore deeper. This feedback is consistent with how managers iterate on a decision by decomposing it into tractable portions. This suggests that representations could support representing a decision at different levels of resolution.

Third, managers wanted different materials to express decisions at different levels of resolution---visual, non-numerical material to represent objectives and tabular calculation structures to hold and process lower-level data after it had been quantified. Yet, this need results in frustrations with copying over or directly expressing higher-level representations in spreadsheets. Therefore, we explore the idea of using distinct but linked materials to represent different decision levels.

Finally, managers valued guided reflections but resisted digital proxies of stakeholders as reflection guides. Their feedback suggests that reflection is best supported via written prompts. We used the preceding reflections as a design guide and prototyped DISCERN.

\section{DISCERN}
Based on our formative work, we developed a technology probe \sys{}, instantiated as an extension for Microsoft Excel. \sys{} has two main functions motivated by our formative work: 
(1) it supports the externalization of multi-level decisions as a tree artifact;
and (2) at each node of the tree, it provides both a visual representation---to support reasoning about objectives---and a linked tabular representation---to hold and reason about data. 
We leveraged this technology probe to challenge our ideas and generate new questions for the design of decision representation tools, rather than committing to a specific design direction that managers may not ultimately desire. We evaluated \sys{} in the context of user enactments with managers (described in Section~\ref{sec:enactment}). In this section, we describe the design of \sys{}, its goals as a technology probe, and its implementation.

\subsection{Design}
\begin{figure*}[t]
    \centering
    \includegraphics[width=\textwidth]{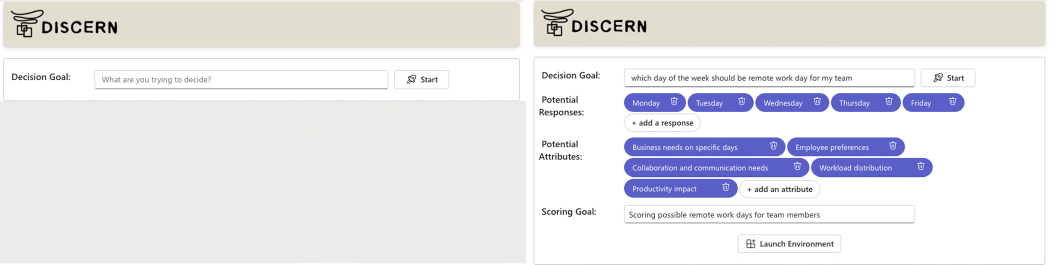}
    \caption{When a user wants to set up a decision within \sys{}, they are asked to provide their decision goal (Left), after which they are asked to provide a list of alternatives (Potential Responses (Right)) and a list of objectives they want to consider in making the decision (Potential Attributes (Right)). On clicking ``launch environment,'' the user is presented with the interface shown in Fig. \ref{systemfig}.}
    \Description{
    “Two screenshots of the DISCERN extension: on the left is the landing page, which contains a textbox labelled ‘Decision Goal:’ containing placeholder text ‘What are you trying to decide?’ A ‘Start’ button sits to the right. On the right, the ‘Decision Goal’ is filled-in with ‘which day of the week should be remote workday for my team.’ Three new labels follow: ‘Potential Responses,’ ‘Potential Attributes,’ and ‘Scoring Goal.’ There are five editable options next to ‘Potential Responses,’ ‘Monday,’ ‘Tuesday,’ ‘Wednesday,’ ‘Thursday,’ and ‘Friday,’ along with an ‘+ add a response’ button. Similarly, editable options next to ‘Potential Attributes’ include ‘Business needs on specific days,’ ‘Employee preferences,’ ‘Collaboration and communication needs,’ ‘Workload distribution,’ ‘Productivity impact,’ and an ‘+ add an attribute’ button. Next to ‘Scoring Goal’ is a textbox with ‘Scoring possible remote workdays for team members.’ At the bottom is a ‘Launch Environment’ button."
    }
    \label{startup}
\end{figure*}
\begin{figure*}[th]
    \centering
    \includegraphics[width=0.94\textwidth]{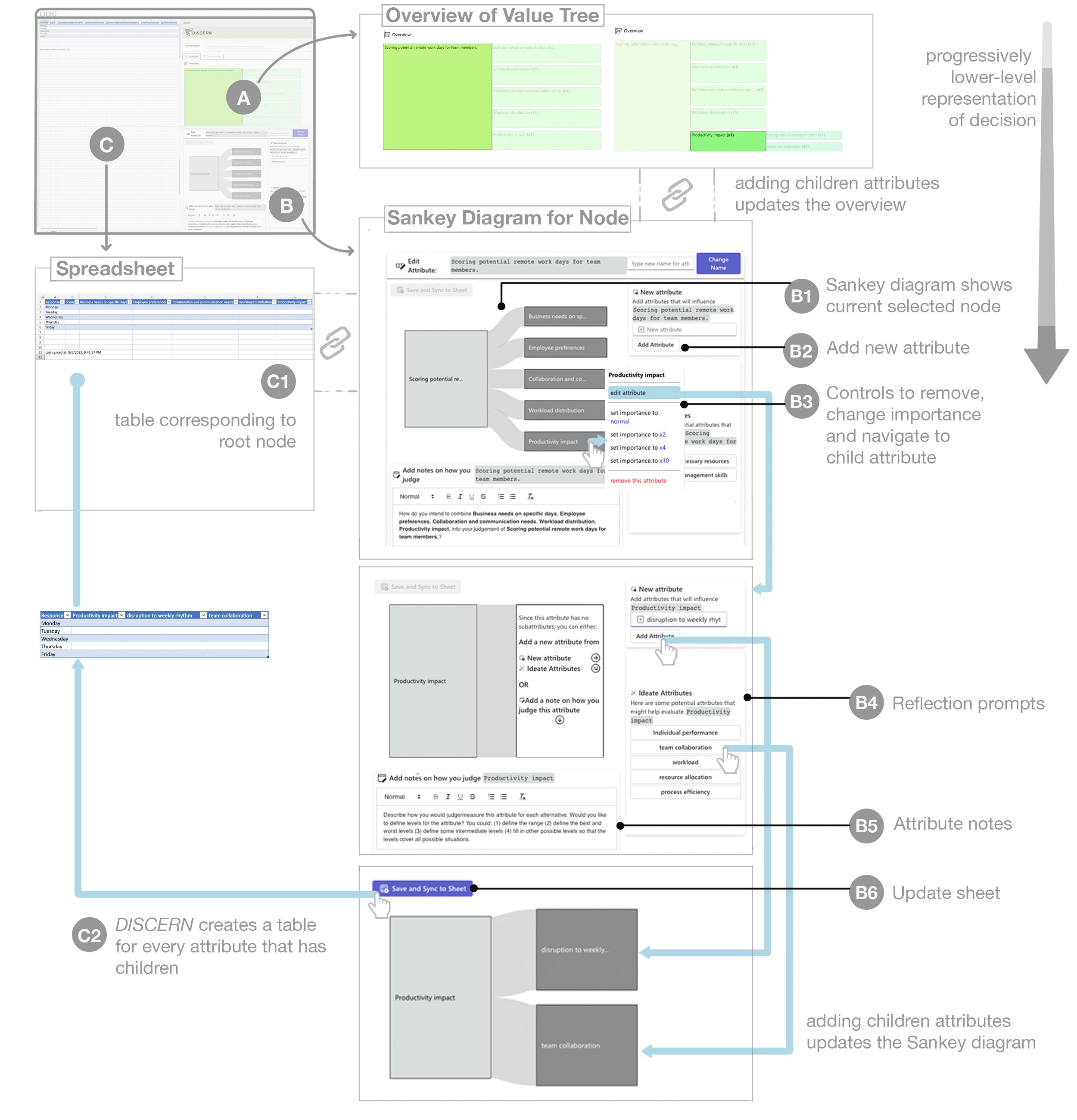}
    \caption{\sys{}'s interface. Panel A shows an overview of the value tree, presented as an Icicle diagram. Users can click on attributes to navigate to them. Panel B allows users to edit the currently selected tree node. The node and its children are presented as a Sankey figure (B1). For a chosen node, users can add new children (B2), change a child's importance level (B3), and leave notes (B5). A list of suggestions reflects potential children that the selected node can be decomposed into (B4). Users can navigate to the selected node's children through the Sankey diagram (B3). Adding children to the selected node updates both the Sankey diagram as well as the value tree overview. Finally, \sys{} also creates spreadsheet tables for each non-primitive node (nodes with children), where users can store raw data (Panel C). These tables are generated and updated automatically when manipulations of the value tree are synced to the sheet (B6). For example, adding children to the ``productivity impact'' node and syncing the update creates a new corresponding table (C2).}
    \Description{
    “Three linked components: (A) Overview of Value Tree, (B) Sankey Diagram for Node, and (C) Spreadsheet. Within the Value Tree, the Overview contains the user-specified decision goal. In this example, the goal is ‘Scoring potential remote workdays for team members.’ In the Sankey Diagram, the goal is represented by the root node and is connected to child nodes representing attributes such as ‘Employee preferences,’ ‘Business needs on specific days,’ ‘Collaboration and communication needs,’ ‘Workload distribution,’ ‘Productivity impact.’ On the left of the diagram is a (B1) pane showing the currently selected node. To the right is a (B2) ‘New attribute’ widget to add a new attribute. Clicking on a child node (e.g. ‘Productivity impact’) brings up a (B3) context menu with controls to remove, change importance, and navigate to the pane for that child attribute. Below the new attribute widget, there is a (B4) ‘Ideate attributes’ widget with reflection prompts to help evaluate the child attribute. For ‘Productivity impact,’ examples include ‘individual performance,’ ‘team collaboration,’ ‘workload,’ ‘resource allocation, and ‘process efficiency.’ Underneath the selected node panel is an (B5) ‘Attribute notes’ widget to ‘add notes on how you judge [the selected attribute].’ Above the selected node panel is a (B6) ‘Save and Sync to Sheet’ button to update the sheet. For each attribute that has a child attribute, a (C2) table is created within the spreadsheet below the (C1) table corresponding to the root node, or highest-level objective. From the Overview of the Value Tree through the Sankey Diagram for Root, Intermediate, and Leaf Nodes, a progressively lower-level representation of decision is reflected."
    }
    \label{systemfig}
\end{figure*}
Within \sys{}, every decision is represented as: (1) a list of alternatives (i.e., what is being evaluated), (2) a value tree (i.e., why and how the evaluation is done), and (3) raw data (i.e., the basis of the evaluation). When a user wants to set up a decision within \sys{}, they are asked to provide their decision goal (Figure \ref{startup} (Left)), after which they are asked to provide a list of alternatives (Potential Responses in Figure \ref{startup} (Right)) and a list of objectives they want to consider in making the decision (Potential Attributes in Figure \ref{startup} (Right)). On clicking ``launch environment,'' the user is presented with the interface shown in Figure \ref{systemfig}, which includes a spreadsheet as well as an interactive side panel. We now outline the main features of \sys{} while also describing how these are presented in the user interface. 

\subsubsection{Supporting multi-resolution representations through a tree artifact}
\sys{} uses the value tree as its primary representation of a decision, consisting of a root node and intermediary and primitive attributes. At the root of the value tree sits the user-specified goal of the decision (e.g., ``deciding which day of the week should be remote work day of the week for my team''). The root node can be further decomposed into a hierarchy of intermediate attributes that represent objectives the decision-maker is explicitly considering. For example, ``business needs,'' ``employee preferences,'' and ``productivity impact'' may represent objectives being considered when deciding the remote work day. Leaf nodes or primitive attributes are intended to directly map to measurements or judgments (e.g., ``number of existing meetings that would be disrupted by working remotely on a given day of the week''). They may be measured quantitatively or judged subjectively but are not derived from other attributes. Moving from the root node to a primitive attribute provides an incrementally lower-level view of the decision and how intermediate attributes (objectives) are mapped to primitive attributes. 
For instance, ``productivity impact'' may be decomposed into ``number of disrupted meetings.'' 
\sys{} allows decisions to be represented at any level of resolution and does not enforce a certain tree depth. 
Thus, by navigating the tree structure, the user can view the objectives being considered, how the objectives are further decomposed, and finally, the primitive objectives that directly map to raw input.

\subsubsection{Different but linked materials to converse with the value tree and raw data}
Our formative work suggested that, while spreadsheets provided a useful tabular representation to hold information and perform quantitative analyses, their ``language of grids'' did not afford managers the granularity to express understanding and clarification of higher-level objectives. On the other hand, more expressive material such as sketches and notes allowed reasoning about higher-level objectives but had to be manually linked to the locations where the data was held. \sys{} attempts to address these complementary drawbacks by providing an alternate visual language for iterating on the value tree while using the ``language of grids'' to store and analyze data. 
\paragraph{Visual language to converse with the value tree:} We use a visual representation for the value tree because managers in our formative study preferred visual approaches to reason about higher-level objectives. \sys{} visualizes the value tree as an Icicle diagram \graycircle{A} which provides a summary of the tree and supports immediate navigation to specific nodes. 
To support constructing and viewing the value tree, we represent a selected node as a Sankey diagram \begin{small}\graycircle{B1}\end{small}. 
The diagram provides controls to add child attributes to the selected node \begin{small}\graycircle{B2}\end{small}, navigate to a child attribute, remove a child attribute, or change the node's importance level \begin{small}\graycircle{B3}\end{small}. 
Because managers wanted the ability to express the importance of different objectives but were hesitant to use precise numbers, \sys{} allows users to express the importance of an attribute (for the evaluation of its parent attribute) in discrete levels (x1, x2, x4, x10) which are encoded visually as heights in the visualization (e.g., the importance of ``employee preferences'' has been set to x1 \begin{small}\graycircle{B1}\end{small}).
\paragraph{Spreadsheets to converse with raw data:}
Because managers were already comfortable using spreadsheets for data recording and analyses, we retained this material for conversing with raw data \graycircle{C}. 
\paragraph{Linked material:} 
\sys{} provides a link between the value tree and the spreadsheet by synchronizing the structure and data.
For instance, if a decision-maker chooses to list out ``employee preferences,'' ``productivity impact,'' and ``business needs'' as attributes important for ``deciding the remote work day,'' they may create structures to store their assessment of each alternative (each day of the week) for each criterion in the spreadsheet. Therefore, \sys{} provides users the ability to automatically create a table for each non-primitive attribute by syncing the value tree to the sheet \begin{small}\graycircle{B6}\end{small}. 
For each non-primitive attribute, this creates a table with the alternatives as rows, the attribute as the first column, and its children as subsequent columns. As shown in Figure \ref{systemfig}, if the user decomposes what was earlier a primitive attribute (``productivity impact'' into ``disruption of weekly rhythm'' and ``team collaboration''), \sys{} creates a new table for the newly decomposed attribute \begin{small}\graycircle{C2}\end{small}. Similarly, updates to a non-primitive attribute, such as the addition or removal of child attributes, sync with the table corresponding to the attribute. 
To provide flexibility for the users to decide which tables they would like to use to record and analyze information, \sys{} presents tabular structures for \textit{all} non-primitive attributes (\begin{small}\graycircle{C1}\end{small} shows a table corresponding to the highest-level objectives). 
\sys{} doesn't create tables for primitive attributes with the expectation that measurements or judgments corresponding to these attributes can be stored in the table for their parent attribute (e.g., judgments of impact on ``team collaboration'' can be stored in the corresponding column of the ``productivity impact'' table \begin{small}\graycircle{C2}\end{small}). We note that \sys{} only manages the structure of tables created by it, and users can continue using the rest of the spreadsheet just as they would normally. Whenever \sys{} adds a new table, it is added beyond the used range of the sheet.

\subsubsection{Reflection and note-taking}
For every attribute in the value tree, \sys{} suggests a list of potential subattributes that the selected attribute can be decomposed into \begin{small}\graycircle{B4}\end{small} (clicking them adds them as children to the currently selected attribute). These were generated by prompting GPT-3.5 with the decision goal and the value tree as the context. Every attribute in the value tree also has a note pane with a prepopulated question for reflection. For non-primitive attributes, it asked the user how they intended to combine their judgments of the attribute's children into their judgment of the attribute (e.g., ``How do you intend to combine `disruption to weekly rhythm' and `team collaboration' into your judgment of `productivity impact'?''). For primitive attributes, it asked the user, ``Describe how you would judge/measure this attribute for each alternative.'' Together, these features support reflection that aligns with managers' feedback in the formative studies---through simple textual prompts.

\subsection{\sys{} as a technology probe} 
Probes are commonly used instruments in HCI research to engage participants early in the design process~\cite{jorke2023pearl, gaver1999design, hohman2019gamut} and to find out about the ``unknown'' when deployed~\cite{hutchinson2003technology}. \textit{Technology} probes are a type of probe that takes the form of functioning technological artifacts and are ``open-ended with respect to use and users should be encouraged to reinterpret them''~\cite{hutchinson2003technology}. Consistent with this, \sys{} does not prescribe a workflow: users can choose to use its representations whenever and however they want to in their decision-making process. Hutchinson et al. ~\cite{hutchinson2003technology} suggest that well-designed technology probes should balance three goals: \textit{social science}: ``understanding the use and the users''; \textit{engineering}: ``field testing the technology''; and \textit{design}: ``inspiring designers and users to think of new kinds of technology to support their needs and desires''.  As a technology probe, \sys{} goals are as follows:
\begin{itemize}
    \item \textit{Social science:} investigating when and how managers want to (1) externalize their decision-making processes and (2) share externalized representations of their decisions.
    \item \textit{Engineering:} evaluating whether a tree representation and a visual language for manipulating it make it easier to reason about higher-level objectives than spreadsheets alone.
    \item \textit{Design:} understanding how \sys{} supported or complicated managers' needs.
\end{itemize}

\subsection{Implementation}
\sys{} is implemented as an extension for Microsoft Excel\footnote{https://www.microsoft.com/en-us/microsoft-365/excel}, using the React\footnote{https://react.dev/} front-end framework, D3.js\footnote{https://d3js.org/}, and the Excel JavaScript APIs\footnote{https://learn.microsoft.com/en-us/office/dev/add-ins/reference/overview/excel-add-ins-reference-overview}. We used the GPT-3.5 API provided through Microsoft's Azure OpenAI Service\footnote{https://azure.microsoft.com/en-us/products/ai-services/openai-service} for the prompts in \begin{small}
    \graycircle{B4}
\end{small}. The value trees constructed by the user, and other edits within \sys{}'s interactive panel, are saved to the spreadsheet being used so that they are persistent across reloads. 

\section{User Enactment Study}\label{sec:enactment}
\label{user-enactment}
Mirroring \sys{}'s goals as a technology probe, this study investigated the following questions:
\begin{itemize}
    \item When and how do managers want to externalize their decision-making process?
    \item When and how do managers want to share externalized representations?
    \item Does \sys{}'s tree representation make it easier to reason about higher-level objectives?
    \item In what ways does \sys{} support or complicate the needs of managers?
\end{itemize}
Because organizational decision-making is inherently a social practice, activating \sys{} as a probe requires situating the use of the artifact in the social context of making a decision. Therefore, we use the method of user enactments to investigate our research questions. In user enactments (UEs), designers ``ask users to enact loosely scripted scenarios involving situations they are familiar with as well as novel technical interventions designed to address these situations''~\cite{odom2012fieldwork}. UEs embrace role-playing and performance as a means to engage users in the exploration of technological futures. We describe the development of our user enactments, using different levels of abstraction (scenario, situation, stuff) in the Experiential Futures Ladder proposed by Candy and Dunagan~\cite{candy2017designing}. We describe the decisions we made at different levels of abstraction in the ladder and our rationale:

\textit{Scenario (specific narrative proposition):} We crafted our scenario around two fixed decisions, for a hypothetical team, in a fictitious organization called `HeliCorp'. Each manager participant had to confer with and make decisions on behalf of their team which comprised four individuals who were themselves role-playing characters assigned to them. Participants were asked to make two decisions for their team: (1) which day of the week should be a remote work day and (2) which hours of the day should be quiet hours. For each decision, we provided managers with a predetermined list of possible courses of action (Monday-Friday for the first decision, and 2-hour blocks of time between 9 a.m. and 6 p.m. for the second decision). We chose to create this fixed decision context for three reasons. First, the context of participants' own real-world team decisions could vary widely, and standardization ensured participants reacted to the same stimuli. Second, asking managers to use \sys{} for their own real-world team decisions could heighten the risk of surfacing potentially sensitive content. Third, by retaining control over the decision and the preferences of the team members, we were able to present participants with synthetic (but realistic) decision scenarios that were relatively subjective, where team members' individual preferences were in conflict and picking a course of action involved tradeoffs between competing objectives. \ed{Prior work suggests that UEs that involve intentionally provocative scenarios can produce rich insights~\cite{odom2012fieldwork}. We chose scenarios that reflected current conversations within the organization about hybrid work to make the UE feel more real. Because both these decisions tend to be intertwined with people's work-life balance preferences, they provide an intentionally provocative scenario to prompt the managers to reflect on relational aspects in decision-making. In keeping with the intention of UEs~\cite{odom2012fieldwork}, these scenarios were not intended to be canonical desicision-making tasks but rather boundary objects to ground speculation with participants.}

\textit{Situation (the circumstances of encounter):} Within our UE, both decisions progressed through two enactment events, each event taking place in a separate session in our study. In the first event, the participant was made aware of both decisions. They were informed that they would have the opportunity to confer with their team during a virtual team meeting---the second event---during the second study session. In the first event, ``planning for a team meeting,'' they were asked to use the artifact provided to them to freely externalize their decision-making process in preparation for the meeting where they would share the structures they create with their team. 
In the second event, ``team meeting,'' the participant was asked to use previously created artifacts to engage in a discussion with the four team members and incorporate their input to arrive at a final decision for each task. The four team members were individuals recruited to play the role of one of eight characters. They were each assigned a unique character based on their subjective assessments of how similar they thought they were to each character during their intake. They were given instructions about how to role-play that character during the discussion.
The instructions included each character's preference for the remote workday and quiet hours as well as a detailed list of rationale for their preferences. In both the first and the second events, activities relating to the two decisions progressed sequentially. In the first event, managers were asked to plan for discussions about the first decision before moving on to planning for the second decision. Similarly, in the second event, managers were asked to conclude discussions about the first decision before moving on to discuss the second decision.

\textit{Stuff (particular artifacts):} One of the goals of our study was to evaluate whether \sys{} made it easier to reflect on higher-level objectives when making decisions. Therefore, we presented participants with two tools during the enactment: Excel with and without \sys{}, one to be used for each of the two decisions. We chose to contrast \sys{} against the traditional Excel since that is what most managers in our formative studies described using. In both sessions of our study, participants were asked to use Excel with \sys{} for the first decision and traditional Excel for the second decision. We counter-balanced which decision (``remote work day'' or ``quiet hours'') was presented first. 

We provide our study materials in the Supplementary Material, which also includes the exact text describing the decisions, the characters assigned to the team members, and the characters' preferences.

\subsection{Study Protocol}
We recruited four managers to participate in our UE; two had participated in the survey (but not the interviews) and two had not participated in previous studies. We recruited 14 individuals to play the role of the team members. They participated only in the second session for each manager (four in each session) and were allowed to participate in more than one manager's second session.

\textbf{Session 1}: We introduced the participants to the enactment scenario and gave them an overview of the \sys{} interface. They were presented with the first decision, prompted to list three key objectives to prime them for the planning enactment, and then asked to plan using \sys{}. This process was repeated for the second decision using Excel instead. The session concluded with a short interview and lasted 60 minutes.

\textbf{Session 2}: Team members received character handouts pre-session and joined a video call with character-specific display names to simulate a realistic virtual team meeting. The manager was briefed separately and told they had to share the artifacts from the previous session but could use them however they wanted. After this, the research team briefed the group collectively, including the manager, to help them ease into the situation. The manager then screenshared the previous day's artifacts, guiding a 15-minute discussion on each decision. The session ended with a short interview of the manager and spanned 60 minutes.

We video-recorded all the sessions. The first and second authors wrote memos during and after the sessions. They met over the course of a month to repeatedly review and discuss the videos and their memos to draw out underlying themes. They also created conceptual models and affinity diagrams to reveal connections across participants.

\subsection{Findings}

\subsubsection{\textbf{\sys{} provided a more expressive medium to iterate on and understand objectives}}
All participants mentioned that they found it easier to express objectives using \sys{} than when just using spreadsheets. Comparing \sys{} to their current practice of using spreadsheets, P2 mentioned: \textit{``I would be working most of my time just to keep configuring the table, which I've done, you know, in the past versus just being able to add that table right in there and and add the attributes.''} P3 specifically appreciated having the visual representations: \textit{``I'm a very visual person so I really like translating those, like, boring tables into something that's more connected and visual, right. That's super helpful for me.''} P4 stated:
\begin{quote}
    \textit{I think this could be a good way to make you sit and think through that rubric for that decision-making up front and not sort of, stumble, you know, stumble into it. So I think it's a good I a good framework and I think \sys{} was easy to use.}
\end{quote}

Participants (P1, P2, and P3) also mentioned how it could help them identify where to pay attention in conversation. P2 contrasted their team discussions across the two decisions: 
\begin{quote}
    \textit{[With only Excel], I was so busy trying to think through like build the table and put the right titles and stuff on it that I kind of cut myself short from being able to get the information I needed. So I think \sys{} was really, I mean it was clear in the conversation. I had more information to go with and a better conversation because of this tool.}
\end{quote}
P1 mentioned how they might \textit{``use it to think through the essential criteria and the questions to ask.''} Similarly, P3 suggested, \textit{``For the thoughtful exercise of putting things together and asking the right questions, it can save a lot of time with the decision-making. And it can help actually facilitate these conversations.''}

Participants saw \sys{} as being especially useful for more \textit{``complex''} decisions (P4). P2 elaborated:
\begin{quote}
    \textit{If I was trying to make a decision with multiple facets to you know what would influence my answer, \sys{} would clearly be easier. I also found it easier for me to think through what were the sub-attributes by having it in there versus trying to kind of on the fly.}
\end{quote}
P4 described how they might turn to \sys{} for \textit{``meaningful, fairly impactful decisions.''} P3 mentioned how managing multiple criteria or inputs from multiple stakeholders could be easier with \sys{}.

\subsubsection{\textbf{Managers did not want to purely calculate nor broadcast calculation}}
Participants varied in the extent to which they wanted to perform calculations for the decisions in the scenario. During the enactment, P2 and P3 directly used the tables they had created to seek inputs from their team. However, P1 and P4 chose to have a free-form conversation, asking open-ended questions to their team. 
During the semi-structured interviews, P1 clarified that they would not want to use formal criteria and scores because they thought \textit{``people's work-life balance is very personal.''}
They suggested they would feel more comfortable using numerical scores for \textit{``technical choices, process engineering, prioritizing, you know, like, work.''}. 
Further, even though P2 used the tables to source input, they recorded X's in the tables instead of numbers explaining to their team how \textit{``the X's are really just a tool for me to try to create some type of heat map. It may not be super evident.''} P3 did record numbers in the table during the meeting but mentioned during the semi-structured interview: 
\begin{quote}
    \textit{I would keep it more private and just kind of let the conversation flow because once you put it out there, people see it as more formal maybe. And especially when they see scoring. It's like ohh, what are you scoring me on? What does that mean? Yeah, it may sound a little bit more serious than it actually is.}
\end{quote}
These findings suggest that the desire to perform and communicate objectivity or quantification may vary across managers and the specific situation.

\subsubsection{\textbf{Fuzzily explicated objectives enabled compromise}}
One noticeable tension that \sys{} and spreadsheet tables created was that it made objectives and the decision-making process seem finalized. Even though it was designed to support iteration, none of its representations visibly communicate a sketch-like nature. Costs borne along attributes that were not listed by managers could then be interpreted by team members as being undervalued or dismissed in the decision. We observed P1, P2, and P4 develop strategies to work around this when they wanted to communicate that the decision-making was malleable and that certain costs were not undervalued. For instance, P2's tables recorded each person's top two preferences for the remote work day, but one character's preferences were such that all but one day were equally undesirable. To address the character's concerns about an unfair compromise, P2 took a moment to explain: 
\begin{quote}
    \textit{There isn't necessarily a double vote or anything like that, but I clearly know where you stand, that Thursday is really the perfect day for you. I have to consider that as part of the overall answer. Does that help?}
\end{quote}

P1 and P4 chose to abandon the tables altogether and instead began by simply asking team members what their constraints were, using free-form notes within the sheet to record responses. P1 mentioned their rationale: \textit{``I'm considering each person independently, so like I would get everyone's like top three [times] and look at the intersection of the top three times.''} P4 adopted a similar strategy, explaining to their team: \textit{``I wanted to reach out to each of you to figure out what the limitations might be, what constraints there might be. And I understand each of your sort of home situations a little bit.''} Doing so could allow each team member to independently communicate what they valued, as opposed to having a seemingly rigid value system imposed on them. It also served to show team members that they might all have to be willing to compromise: \textit{``We know we're not gonna be able to please everybody. So we need to come to, you know, everybody's gonna have to give a little bit.''} (P1). By allowing team members to hear each other's full set of constraints, P4 was also able to encourage them to be accommodating: \textit{``After hearing everybody and considering their input, maybe you can just pop in the chat here your preferred day or options, and it doesn't have to be one.''} Such conversations can be prohibited when what matters seems solidified from the get-go. This suggests that decision representations may be more apt for supporting consensus-building if they allow the facilitator to control the level of fuzziness they communicate.

\subsubsection{\textbf{\sys{}'s top-down construction clashed with bottom-up reasoning}}
The divergence in P2's and P3's decision-making from that of P1's and P4's also revealed how \sys{}'s top-down construction of the value tree could clash with attempts to make decisions in a more bottom-up manner. \sys{}, by design, promoted a focus on objectives: It asked managers to articulate what they thought was important as a way to begin differentiating between choice alternatives.  P1 and P4, however, had a more bottom-up approach. They wanted to first understand the dimensions along which the choice alternatives varied: what choice alternatives would definitely not work and what would be lost in moving from one to the other. Decision-making progressed by eliminating alternatives: \textit{``I think we have a consensus that you know, non-Friday''} (P1). P1 further mentioned how they tried to begin by finding an overlap in choices: \textit{``I'm trying to like phrase the question so that I can get some sort of overlap answer. So this is I'm establishing sort of the constraints on the negative side.''} Similarly, P4 articulated to their team, ``My goal here is to consider everything that you've shared with me and the constraints and whether we have some workarounds''. This provides another reason why they abandoned the tables in the spreadsheet in favor of open-ended questions. On the other hand, P2 and P3's articulation of the decision aligned more with a top-down representation: \textit{``So those are the aspects that we're going to cover here. So I'm just going to go around the room and get input from everybody''} (P3). They used previously decided objectives to steer towards and away from certain options: \textit{``Clearly, if you look at where the X's are from the least productive standpoint or when you have the most conflicts with having quiet time, it's in the afternoon for all of us, right?''} (P2). This suggests to us that a strictly top-down or a strictly bottom-up approach to constructing decision representations might not accommodate the diverse ways in which decisions are actually made.

\section{Discussion}
Our formative studies (Section \ref{formative}) revealed how managers did not want tools that obviate their face-to-face interactions with stakeholders but instead wanted representations to simply hold and reason about what they had heard. Based on our findings, we developed \sys{} which we used as a technology probe in our user enactment study (Section \ref{user-enactment}). This further revealed how managers wanted tools that would allow them to qualculate and express objectives
fuzzily to enable consensus. 

\camera{Our work makes an empirical contribution to HCI literature on digitally-supported social decision-making by providing an account of social decision-making practices that are more intimate and less formalized than those previously studied. In particular, our studies describe the relational considerations underpinning decision-making practices of line managers that we suggest are likely to be present in other social decision-making contexts where the maintenance of relationships is paramount. Our work also makes an empirical contribution to literature on information management and visual analytics in the workplace by presenting the unmet communication and analysis support needs of line managers. Here, our work provides one more piece of evidence that current tools are not flexible enough to support the way decision-making progresses~\cite{dimara2021unmet}, iteratively towards higher levels of fidelity.}

\camera{In the rest of this section, we interpret our findings in the context of prior work to contribute to design knowledge in digitally-supported social decision-making. Specifically, we consider how our observations reveal the limits of common premises underlying current tools (Section \ref{priorwork}), about how decision-making can be supported. Based on this, we contribute implications for the design of digital supports for social decision-making. We also discuss opportunities for future research and practice that aims to develop decision support tools for social decision-making, and especially in workplaces.}

\ed{\subsection{Making social decision-making, social: from an informational to an interactional perspective}
Digitalization of work-related activities has led to commercial and research interest in how data can be leveraged to speed up workplace decision-making~\cite{tursunbayeva2022ethics}.
However, our observations suggest that the current focus on primarily analytics and forecasting may be incongruent with the decision-making needs of line-managers. Our research revealed that participants did not view data as free-standing. Participants reported that stakeholder interactions were necessary to contextualize data, draw on others' expertise, and understand preferences. These conversations, and reflections on them, often formed a significant part of managers' decision-making activities. These observations validate previous calls to expand the scope of decision support tools beyond data analysis to tools for interactional decision-making as a greater priority~\cite{dimara2021unmet}. Furthermore, participants did not want to listen via digital surveys or preference elicitation tools. Instead, they wanted representations that recorded understanding from face-to-face interactions. Participants described how they used the elicitation process to infer judgments beyond what was said. They also thought face-to-face listening was important for making people feel heard: interpersonal interactions facilitated trust and compromise.

Our observations suggest that digitally mediating stakeholder interactions may not be suitable in more intimate social decision-making. When decision-makers are socially detached from affected stakeholders, and where direct interaction may be infeasible, it may be necessary to convey stakeholders perspectives digitally, as information. We suggest that when decision-makers must interact more closely with affected stakeholders, it is better to view stakeholder interaction less as information trasmitted through a conduit~\cite{reddy1979conduit} and more as socially enacted phenomena~\cite{boehner2005affect}. We find an elaboration in Suchman's book \textit{Plans and Situated Actions}~\cite{suchman1987plans}, which has been influential in CSCW research: ``Conversation is not so much an alternating series of actions and reactions between individuals as it is a joint action accomplished through the participants' continuous engagement in speaking \textit{and} listening'' (emphasis in original). This suggests that stakeholder's dynamic perspectives aren't merely data to be extracted and transmitted, but that those perspectives and their meaning are socially enacted: what someone wants and what that entails for the decision is co-constructed through dialogue. 

Approaches to digitally elicit preferences and numerically represent values, largely draw on an information theoretic view of communication, inspired by cognitive science~\cite{reddy1979conduit}. We suggest that the design of decision supports can also benefit from a phenomenological viewpoint~\cite{harrison2007three}, considering specifically, as Boehner et al. suggest, a ``move from technologies of representation to technologies of participation''~\cite{boehner2005affect}. As an implication of this perspective for design, we suggest that rather than enforcing a ``correct'' representation of decision-making on the decision-maker, tools can serve as platforms with flexibility for decision-makers to enact their social decision-making practices.

\subsection{Designing representations to anchor group deliberation}
Instructional material for managers~\cite{alemi2007decision, watson1987decision}, decision support tools~\cite{gratzl2013lineup, pajer2016weightlifter, carenini2004valuecharts}, and even professional decision analysts~\cite{cabantous2010decision, howard1966decision} assume that decision-makers should select among several options by scoring them, through individual choice or group deliberation. This assumption underlies prior work's premise that numerically representing values can help decision-makers better clarify their values. However, we noticed that such representations if used for group deliberations, could disrupt consensus seeking by making individuals more attuned to who might gain and lose from each option. This finding has also been previously reported~\cite{kraemer1988computer}. Building on our observation, we suggest that representations intended to anchor deliberative conversations can benefit from a layer of ambiguity or fuzziness, rather than pinpointing values and disagreement numerically. For instance, interfaces that enable communicating ambiguity or flexibility in value representations is an interesting direction for future exploration. 

\subsection{What does qualculation mean for the design of decision support tools?}
We found that managers practice qualculation in their decision-making. This finding suggests that although tools that numerically represent values are useful, the level of numerical analysis ought to be situational: what constitutes appropriate action depends on what is accountably justified in the eyes of those experiencing the situation. An individual deciding where to host a family dinner may not want a tool that supports precise calculation of costs and benefits but may still desire some level of support in comparing nearby restaurants. Similarly, decisions within organizations exist at different levels of formalization. A primary focus on supporting highly formalized decisions and precise mathematical modeling may overlook the organizational decisions that do not require that level of analysis but which may still benefit from technological support. Participants in our formative studies also highlighted how each decision progressed iteratively through different levels of formalization, yet current tools only permitted polar levels of analysis: \textit{“You can either build that tree or have the data”} (P6). Recent research studying the needs of organizational decision-makers arrived at a similar conclusion: ``Survey and interviews suggested that decision makers lack `decision making' tools designed to support the flow along all decision phases''~\cite{dimara2021unmet}. Therefore, future research should focus on flexible interfaces~\cite{dimara2019interaction, dimara2021unmet} that support the progression of decision-making across levels of formalization. \sys{} presents one vision of supporting intermediate levels of analysis. However, as our user enactments revealed, \sys{} still has limitations; it may not adequately support early-stage brainstorming. We discuss this and other design opportunities next.

\subsection{Future Work}
Our findings highlight design opportunities for tools that support decision-making within organizations. Here we detail two of these opportunities.
\subsubsection{Decision-making as design}
In early stages of a decision, groups might benefit less from structure and more from tools that support brainstorming. That is, decisions in early stages may be better represented as design problems~\cite{simon1996social}. Our user enactments revealed how managers often abandoned \sys{}'s representations to fuzzily explicate their objectives to seem less final. One possible reason is that, even though \sys{} provided ways to articulate structure at different levels of abstraction, such structure imposed rigidity. In the early stages, managers were still understanding the problem and did not want to impose structure on the group. This suggests opportunities for creating discourse platforms where structure can be invoked only when the need is felt. Furthermore, our work focused on decisions where the options were known apriori; these were defined once in \sys{} and then left fixed. In many real-world settings, managers are also tasked with synthesizing alternative options for their teams. Future work could investigate how tools can help decision-makers ideate alternate courses of action. Drawing on a growing body of work that uses Large Language Models to envision futures~\cite{buccinca2023aha, park2022social}, one approach could be to propose how stated options may be improved, using the values articulated by the decision-maker and the options currently under consideration. 

\subsubsection{Decision-making as experimentation}
We found that managers' decision-making processes were iterative, resisting delimitation. In some cases, participants described how team members' commitment to an option was conditional on a retrospective evaluation in the near future. Decision support tools, insofar as they support externalizing a decision's initial rationale, can also support retrospective evaluations. In smaller groups where the cost of trying a course of action may be low, the process of choice can start to resemble an experiment: Instead of anticipating future outcomes to pick a single option, a team can experience and reject ineffective options until they arrive at the one that works for them. Tools like \sys{} only incidentally support retrospective evaluation of decisions. There is an opportunity to design tools that more directly support teams in setting up and running their own decision-making experiments, drawing on social computing work on community-led experimentation~\cite{matias2018civilservant, pandey2021galileo}.
}

\section{Limitations}
\ed{Studying the practices of line managers within a technology organization provided us the opportunity to investigate how the first line of management---a large portion of decision-makers within organizations---envisions decision support. However, this context brings three important limitations to this account that future work can try to address. First, our focus on line managers meant that workers' perspectives were only represented incidentally, through the lens of line managers, who are workers of the organization themselves. Future work should center the concerns of workers through mutli-stakeholder co-design processes~\cite{holten2021can, hsieh2023co}. Second, the needs and challenges of line managers in a technology organization are unlikely to represent the challenges of line managers in other kinds of work, such as service work~\cite{spektor2023charting}, and we caution against generalizing these insights to other contexts. Finally, our participant pool comprised line managers in US-based locations, and their perspectives should be interpreted within this cultural context.}

\ed{Design methods---and all research methods---include trade-offs in the kinds of evidence they generate and the kinds of conclusions they facilitate.} Designing \sys{} as a technology probe and investigating its use through a user enactment also involved these trade-offs. For instance, we could not investigate whether or when managers would have preferred to externalize decisions in their own real-world decision-making. \ed{As with other design research, our design activities and choice of user enactment scenarios were intended for rich qualitative accounts rather than statistical validity~\cite{zimmerman2007research}. Other researchers working with the same problem framing may create other artifacts or pursue different design activities, arriving at other, equally relevant conclusions~\cite{zimmerman2007research}. While our choice of design methods does not allow for statistical validity, we believe our work offers opportunities for what Zimmerman et al. describe as ``extensibility''~\cite{zimmerman2007research}: that future attempts to develop tools for social decision-making, especially within the workplace, can build on the knowledge created by our artifacts}. 

\section{Conclusion}
\ed{While current conversations on supporting workplace decision-making tend to focus on data analytics and predictive models, this paper recenters an alternate space where information technologies can play a role: in interactional decision-making. We investigated opportunities for supporting the decision-making practices of line managers. They form the first level of management, and are a significant population of understudied organizational decision-makers, who must make complex workplace decisions while maintaining their relationships with those impacted by their decisions. We presented findings from our iterative design process. In formative work, we found that line managers did not want tools that obviated their listening practices but instead, wanted representations to simply hold their reasoning. Based on findings from our formative study, we developed \sys{}, a technology probe that allows managers to externalize their decision-making. We conducted user enactments where managers were invited to use \sys{}. This further revealed that managers wanted tools that could flexibly enable intuitive and calculative decision-making to support decisions with varying levels of formalization. We observed how publicly quantifying the ramifications of decisions could threaten consensus-seeking in some cases. Based on these findings, we discussed future directions for the design of workplace decision-support tools while also discussing broader implications for designing social decision-making supports. More broadly, we hope that, going forward, this work encourages a greater focus on interactional decision-making in designing technological supports for workplace decision-making.}

\begin{acks}
This work was done while Pranav Khadpe was an intern at Microsoft Research, working with the Human Understanding and Empathy Group  and Viva Insights. We would like to thank members of both teams for the invaluable (and repeated) assistance that made this research possible, and helped us be responsive to what we were learning in the field. In particular, we would like to thank Gonzalo Ramos and Mary Czerwinski for the many discussions and study pilots that helped shape this work. We would also like to thank Chinmay Kulkarni, Geoff Kaufman, and our anonymous reviewers for their feedback, questions, and fresh perspectives that enhanced the depth and overall quality of this work. We thank the managers who took part in our study activities for sharing their insights and time with us, and the interns at Microsoft who participated in our user enactments for their excellent role-playing. We are grateful to Koustuv Saha for bringing the research team together, noticing our complementary interests. Pranav Khadpe would like to thank Kimi Wenzel for help with articulating this work in the 150 words of the abstract. Finally, Pranav Khadpe would like to thank Anna Kawakami, Marc Mascarenhas, Alicia DeVrio, Nathan DeVrio, Conor Lawless, and Alice Zhang for the many ways in which they supported this work that escape precise specification. 
\end{acks}

\bibliographystyle{ACM-Reference-Format}
\bibliography{bibliography}

\end{document}